\documentclass[aps,prb,twocolumn,longbibliography,floatfix]{revtex4-1}
\usepackage{graphicx}
\usepackage{bm}
\usepackage{color}

\begin{document}
\title{The Dynamics of Magnetism in Fe-Cr Alloys with Cr Clustering}
\author{Jacob B. J. Chapman}
\email{Corresponding author: Jacob.Chapman@ukaea.uk}
\affiliation{UK Atomic Energy Authority, Culham Science Centre, Oxfordshire, OX14 3DB, United Kingdom}
\author{Pui-Wai Ma}
\affiliation{UK Atomic Energy Authority, Culham Science Centre, Oxfordshire, OX14 3DB, United Kingdom}
\author{S. L. Dudarev}
\affiliation{UK Atomic Energy Authority, Culham Science Centre, Oxfordshire, OX14 3DB, United Kingdom}

\begin{abstract}
The dynamics of magnetic moments in iron-chromium alloys with different levels of Cr clustering show unusual features resulting from the fact that even in a perfect body-centred cubic structure, magnetic moments experience geometric magnetic frustration resembling that of a spin glass. Due to the long range exchange coupling and configuration randomness, magnetic moments of Cr solutes remain non-collinear at all temperatures. To characterise magnetic properties of Fe-Cr alloys, we explore the temperature dependence of magnetisation, susceptibility, Curie temperature and spin-spin correlations with spatial resolution. The static and dynamic magnetic properties are correlated with the microstructure of Fe-Cr, where magnetisation and susceptibility are determined by the size of Cr precipitates at nominal Cr concentrations. The Curie temperature is always maximised when the solute concentration of Cr in the $\alpha$ phase is close to 5 to 6 at.\%, and the susceptibility of Fe atoms is always enhanced at the boundary between a precipitate and solid solution. Interaction between Cr and Fe stimulates magnetic disorder, lowering the effective Curie temperature. Dynamic simulation of evolution of magnetic correlations shows that the spin-spin relaxation time in Fe-Cr alloys is in the 20 to 40 ps range.
\end{abstract}
\maketitle

\section{Introduction}  
Iron and chromium are the main components of steels. They are both magnetic transition metals. Fe-Cr alloys adopt various structural and magnetic phases depending on the chemical composition and temperature. Theoretical understanding of the correlation between magnetic properties and microstructure provides valuable insight for tuning and characterising the mechanical properties of Fe-Cr alloys and steels. Understanding the coupled mechanical and magnetic properties that evolve as the underlying microstructure changes is particularly significant for steels development, including reduced activation ferritic martensitic (RAFM) steels for advanced fission and fusion reactors \cite{Stork_FusEngDes_2014}. 

Magnetic ordering strongly affects structural and mechanical properties of Fe-Cr alloys. For example, the total free energy resulting from both magnon and phonon excitations stabilises the ferromagnetic (FM) body-centred cubic (BCC) phase of Fe in comparison with the hexagonal close-packed phase structure stabilised by the $s$ and $d$ orbital occupation alone \cite{Hasegawa_PRL_1983,Ma_PRB_2017}. Self-interstitial atom (SIA) defects in all the non-magnetic transition metals form $\langle 111 \rangle$ crowdion configurations \cite{Nguyen-Manh_PRB_2006,Derlet_PRB_2007,Ma_PRM_2019}. On the other hand, SIAs in magnetic Fe and Fe-Cr alloys prefer $\langle 110 \rangle$ dumbbell configurations with antiferromagnetic (AFM) ordering at the core of the defect \cite{Fu_PRL_2004,Olsson_PRB_2007}. On the macroscopic scale, the softening of elastic constant $C'$ near the Curie temperature $T_C$ results from magnetic excitations \cite{Dever_JAP_1972,Hasegawa_JPhysF_1985}.

Magnetic phases of Fe-Cr alloys are fairly complex. The magnetic ground state of pure BCC Cr is an incommensurate spin density wave (SDW) with the N\'{e}el temperature $T_N^{\text{exp}}\approx 310$K \cite{Fawcett_RMP_1988}. Magnetic moments are in antiparallel alignment with magnitudes varying sinusoidally and being incommensurate with the lattice periodicity \cite{Overhauser_PR_1962,Fawcett_RMP_1988,HafnerPRB2002}. The introduction of isolated Fe atoms into the Cr matrix suppress the ordering by destabilising the correlation between electron-hole pairs, which makes the SDW commensurate at 96-98 at.\% Cr \cite{Burke_JPF_1983_b}. Between 81 and 84 at.\% Cr, Fe magnetic moments prevent long range AFM-like correlations, causing the moments to freeze into a spin glass state \cite{Burke_JPF_1983}. For higher concentrations of Fe, Fe-Cr systems attain FM ordering. The Curie temperature $T_C$ is found to be maximised at $T_C^{\text{exp}}\approx 1049$K with Cr concentration near 6 at.\% \cite{Lavrentiev_JPhysCM_2012}. An ordered structure of the alloy at 6.25 at.\% Cr is stabilised by a minimum of the spin-down density of states at the Fermi energy \cite{Nguyen-Manh_ComputMaterSci_2010}. 

Fe-Cr alloys generally adopt a BCC structure across the whole chemical composition range, except at low Cr concentration and at high temperature, where they exhibit an FCC $\gamma$-loop in the phase diagram \cite{Andersson_CALPHAD_1987}. Despite having similar lattice constants, where $a^{Fe}$=2.87{\AA} and $a^{Cr}$=2.91\AA, they do not always form a solid solution. Between room temperature and 700K, a miscibility gap is observed starting from 5-10 at.\% Cr and terminating at 90-95 at.\% Cr. The precise boundaries remain contested \cite{Xiong_CritRevSSMS_2010,Andersson_CALPHAD_1987,Bonny_SM_2008}. Within the miscibility gap, there is a domain of spinodal decomposition. At temperatures above 1100K, iron and chromium are fully miscible. Fe-Cr alloys exhibit anomalous formation enthalpy as a function of chemical composition and temperature \cite{Klaver_PRB_2006,Olsson_JNM_2003}. At low temperature, when Cr concentration is below approximately 10 at.\%, short-range ordering of the Cr magnetic moments gives rise to a negative formation enthalpy, which favours complete dissolution of Cr.

The 475$^{\circ}$C embrittlement in steels \cite{Grobner1973} is a well known phenomenon associated with spinodal decomposition. Thermally aged Fe-Cr alloys undergo slow coherent phase separation into Cr-rich $\alpha'$ and Fe-rich $\alpha$ precipitates if the chemical composition is in the miscibility gap \cite{Grobner1973}. Ion irradiation experiments \cite{Hardie_JNucMat_2013} have shown that irradiation-accelerated Cr precipitation is closely related to the hardening of Fe-Cr alloys. The first wall of fusion reactors is expected to sustain high irradiation doses approaching 10 to 20 displacements per atom (dpa) per full operation year \cite{Stork_JNucMat_2014,Stork_FusEngDes_2014}. 

In general, precipitation during thermal ageing occurs across a time scale exceeding $10^4$ hours \cite{Novy_JNucMat_2009}. Under irradiation, this can be accelerated by 6 to 7 orders of magnitude due to a substantial increase in point defect concentrations enabling fast mass diffusion \cite{Soisson_AM_2016}. 
Despite different time scales, the growth morphology of Cr clusters is indistinguishable between irradiation and annealing \cite{Soisson_AM_2016}, with the exception in the vicinity of sinks, such as cavities and grain boundaries, where precipitate free zones form \cite{Little_JNucMat_1979,Bachhav_JNucMat_2014}.

The kinetics of Cr clustering in Fe-Cr alloys has been studied extensively  through experimental \cite{Furusaka_JPhysSocJpn_1986,Novy_JNucMat_2009,Bley_AMM_1992,Dubiel_JAlloyCompd_2015,Miller_ActaMetall_1995,Kuronen_PRB_2015} and theoretical approaches \cite{Soisson_AM_2016,Kuronen_PRB_2015,Dopico_AM_2015}. However, the effect of Cr precipitation on the magnetic properties of Fe-Cr alloys has largely been ignored from quantitative analysis. This motivates a systematic investigation on the correlation between the magnetic properties of Fe-Cr alloys and their microstructure. 

In this paper we quantitatively explore the relationship of magnetic properties of Fe-Cr alloys with the underlying atomic ordering. We show that the Curie temperature is not only dependent upon the nominal concentration of solute atoms but also on the degree of Cr clustering. Furthermore, using correlation functions we show that the local Curie and N\'{e}el temperatures change in the vicinity of the cluster interfaces. We shall understand these phenomena from the scientific perspective, showing that some of the peculiar magnetic properties stem from geometric magnetic frustration. The randomness of exchange coupling between magnetic moments in Fe-Cr alloys essentially form spin glasses. Knowing the relationship between magnetic properties and precipitation in a quantitative manner may help in the development of a non-invasive engineering technique to estimate the level of Cr precipitation by probing local magnetic properties. 

Our paper is structured as follows. In Section \ref{Sec:method}, the method of Langevin spin dynamics, the Hamiltonian and the quantitative properties of interest are introduced. In Section~\ref{sec:ss}, magnetisation of solid solutions of Fe-Cr alloys are investigated, which are compared to alloys consisting of ideal and impure Cr clusters in Sections \ref{sec:ideal_cluster} and \ref{sec:clusters}, respectively. We examine the change of Curie temperature $T_C$ as a function of the size and chemical composition of clusters. The dynamics of magnetisation characterised using correlation functions of the idealised clusters is examined in Section \ref{ss:time} and for (001) Fe/Cr interfaces in Section \ref{sec:interface}. Our results are summarised in Section \ref{sec:conclusion}. 

\section{Methodology}\label{Sec:method}
\subsection{Spin Dynamics}

Spin Dynamics (SD) is an indispensable tool for studying magnetic excitations. Originating from the Landau-Lifshitz and Gilbert equations \cite{Landau_PZS_1935,Gilbert_TransMagn_2004}, SD has vastly extended our understanding of solitons and magnons in magnetic media  \cite{Tao_PRL_2005,Ma_PhilMag_2009,Ma_PRB_2017}. Recent developments consider both the direction and magnitude of an magnetic moment (or atomic spin) as a three dimensional vector \cite{Ma_PRB_2012,Ma_PRB_2017}. Transverse and longitudinal fluctuations are treated within a unified framework, corresponding to the localised and itinerant property of hybridised $s$ and $d$ electrons \cite{Hubbard_PRB_1979}.

The temperature of a dynamic spin system can be controlled by Langevin thermostat \cite{Brown_PR_1963, Garcia-Palacios_PRB_1998,Ma_PRE_2010,Ma_PRB_2012}. Considering an arbitrary spin Hamiltonian $\mathcal{H}$, the Langevin equation of motion of an atomic spin $\mathbf{S}_i$ can be written as \cite{Ma_PRB_2012}:
\begin{equation}
\frac{d\mathbf{S}_i}{dt}=\frac{1}{\hbar}\big(\mathbf{S}_i\times \mathbf{H}_i\big)+\gamma_s \mathbf{H}_i+\bm{\xi}_i
\label{eqn:sd}
\end{equation}
where the $\mathbf{H}_i=-\partial\mathcal{H}/\partial\mathbf{S}_i$ is the effective field for spin $i$, $\gamma_s$ is a damping parameter and $\bm{\xi}_i$ is $\delta$-correlated random noise, which satisfies $\langle\bm{\xi}_i(t)\rangle=0$ and $\langle\xi_{i\alpha}(t)\xi_{j\beta}(t')\rangle=\mu_s\delta_{ij}\delta_{\alpha\beta}\delta(t-t')$. Greek alphabet subscripts represent the Cartesian coordinates. According to  the fluctuation-dissipation theorem (FDT) \cite{Chandrasekhar_RevModPhys_1943,Kubo_RepProgPhys_1966}, one can obtain a relationship between the fluctuation and dissipative forces \cite{Ma_PRB_2011,Ma_PRB_2012}, namely $\mu_s = 2\gamma_s k_B T$. The magnetic moment and atomic spin are simply related by 
$\mathbf{M}_i=-g\mu_B\mathbf{S}_i$, where $g$ is the electronic g-factor and $\mu_B$ is the Bohr magneton.

We perform SD simulations using our code SPILADY \cite{Spilady}. The source code has been generalised to handle multiple elemental species and complex Hamiltonian forms. In all the calculations, the equations of motion of spins are integrated using the parallelised symplectic Suzuki-Trotter decomposition algorithm \cite{Ma_PRE_2009} with a time step of 1 fs.

\subsection{Magnetic Cluster Expansion Hamiltonian}
We adopt a Heisenberg-Landau type magnetic cluster expansion (MCE) Hamiltonian developed for Fe$_{1-x}$Cr$_{x}$ alloys, which has been used in various Monte Carlo (MC) simulations \cite{Lavrentiev_JNucMat_2009,Lavrentiev_CompMatSci_2010,Lavrentiev_PRB_2010,Nguyen-Manh_JMaterSci_2012}. Terms in the MCE Hamiltonian, contributing to the effective field, can roughly be decomposed into inter-site interaction Heisenberg terms $\mathcal{H}_H$ and on-site Landau terms $\mathcal{H}_L$: 
\begin{equation}
\mathcal{H}=\mathcal{H}_H+\mathcal{H}_L,
\label{eqn:mce}
\end{equation}
where
\begin{equation}
\mathcal{H}_H=\displaystyle \sum _{i \neq j} \bigg( J^{(0)}_{ij} +J^{(1)}_{ij}(\sigma_i+\sigma_j) + J^{(2)}_{ij}\sigma_i\sigma_j\bigg)\mathbf{M}_i\cdot \mathbf{M}_j,
\label{eqn:mce_H}
\end{equation}
and
\begin{eqnarray}
\mathcal{H}_L &=& \displaystyle \sum_{i}\bigg(A^{(0)}+A^{(1)}\sigma_i+\displaystyle \sum_{j \neq i} A^{(2)}_{ij}\sigma_i \sigma_j\bigg) \mathbf{M}_i^2 \nonumber \\
&+& \displaystyle \sum_i \bigg(B^{(0)}+B^{(1)}\sigma_i+\displaystyle \sum_{j \neq i} B_{ij}^{(2)}\sigma_i \sigma_j \bigg)\mathbf{M}_i^4.
\label{eqn:mce_L}
\end{eqnarray}

The magnetic energy of an alloy configuration depends on atomic site occupational variables $\sigma_i$ and the non-collinear magnetic moments of each atom $\mathbf{M}_i$. The discrete occupational site variable $\sigma _i$ takes a value of $\sigma_i=+1$ if the $i^{\text{th}}$ atomic site is occupied by an Fe atom, or $\sigma_i=-1$ if it is an Cr atom.

Since the magnetic interactions depend on the atomic configuration, both the magnetic order and magnitudes \cite{Lavrentiev_PRB_2011} evolve according to the local atomic environment via the Landau coefficients $A_i$, $B_i$ and the Heisenberg exchange coupling parameter $J_{ij}$. The $J_{ij}$ are parameterised up to the 5$^{\text{th}}$ nearest neighbour. $A^{(2)}_{ij}$ and $B^{(2)}_{ij}$ are functions of atomic species in the 1$^{\text{st}}$ and 2$^{\text{nd}}$ coordination shells. The numerical values of the parameters for the Fe$_{1-x}$Cr$_{x}$ Hamiltonian in Eq. \ref{eqn:mce} to \ref{eqn:mce_L} are given in Table 1 of Supplementary Material. Details on the parametrisation are given in Refs. \onlinecite{Lavrentiev_JNucMat_2009,Lavrentiev_PRB_2010}.

Magnetocrystalline anisotropic energy is not included in the MCE Hamiltonian, but can in principle be incorporated phenomenologically \cite{Perera_prb_2016}. It would require extra fitting of parameters from \textit{ab initio} calculations that consider spin-orbit coupling. As we are working on perfect bcc configurations, it is expected that the anisotropic energy is in the $\mu$eV/atom level \cite{Halilov_prb_1998,Xie_prb_2004}. As we will see in the following sections, the origin of magnetic frustration is due to the long range exchange coupling which is in meV level. We may safely ignore the magnetocrystalline anisotropy effects in this study. 

We verified our implementation by performing time averaged benchmarking comparisons for pure Fe, Fe-Cr alloys and (001) Fe/Cr superlattices. They show perfect agreement with MC calculations performed by Lavrentiev \textit{et al.} \cite{Lavrentiev_JNucMat_2009,Lavrentiev_CompMatSci_2010,Lavrentiev_PRB_2010}. The results and details can be found in Supplementary Materials. 

A fundamental difference between Langevin SD (Eq. \ref{eqn:sd}) and MC methods is the process of attaining equilibrium. Whilst both approaches should give the same thermodynamically averaged quantities in the limit of time going to infinity, dynamics simulations show if a system can really achieve equilibrium within realistic timescales. MC methods, on the other hand, do not contain an objective concept of time. It provides no information on the dynamics or time relaxation of events. 

The current model is also applicable to the study of spin glasses. Fe-Cr adopts a spin glass magnetic state at high Cr concentrations \cite{Burke_JPF_1983}. A common theoretical approach is to consider a phenomenological Ising model. In these cases the Heisenberg parameters are drawn from a Gaussian distribution about a non-zero mean \cite{Sherringdon_JPhysF_1975}. On the other hand, the MCE Hamiltonian (Eq.~\ref{eqn:mce}-\ref{eqn:mce_L}) encodes configurational disorder using physically meaningful exchange parameters that were fitted to \textit{ab initio} data \cite{Lavrentiev_PRB_2010}.

\subsection{Simulation Metrics}

An important macroscopic order parameter is the time averaged magnetic moment. The contribution by species $\eta=$ Fe or Cr is calculated using the magnetic moments in thermal equilibrium at a particular temperature $T$ over the trajectory in a time period of $\tau$, assuming ergodicity:
\begin{equation}
\langle \mathbf{M}_\eta \rangle = \frac{1}{\tau}\int_0^{\tau} \left( \frac{1}{N_\eta} \sum _i ^{N_\eta} \mathbf{M}_i(t) \right) dt.
\end{equation}
The summation spans all atoms $N_\eta$ of the species. The averaged system magnetic moment is the contribution from all element types. Magnetisation is the magnitude of the averaged magnetic moment i.e. $\bar{M}_{\eta} = |\langle \mathbf{M}_\eta \rangle|$. 

The susceptibility of an observable with respect to its conjugate field can be calculated from the fluctuations of the observable within a canonical ensemble. This facilitates the calculation of the magnetic susceptibility $\chi$ through the spin dynamics calculations:
\begin{equation}
\label{eqn:sus}
\chi = \mu_0\frac{\langle \mathbf{M}^2\rangle - \langle \mathbf{M} \rangle ^2}{k_BT}, 
\end{equation}
where $\mu_0$ is the vacuum permeability and $\mathbf{M}$ is the averaged magnetic moment within a defined volume.

One can estimate the fundamental timescales of magnetic excitations using the correlation of fluctuating magnetic moments. The correlation of a magnetic moment at site $i$ over a time interval of $\tau$ from a starting time $t$ can be calculated using a time-displaced spin-spin autocorrelation function. 
\begin{equation}
c_0(\mathbf{r},\tau)=\frac{1}{N}\displaystyle \sum_{n}^{N}\bigg\langle \mathbf{e}_i(\mathbf{r},t+\tau)\cdot \mathbf{e}_i(\mathbf{r},t) \bigg\rangle, \label{eqn:rracf}
\end{equation}
where $\mathbf{e}_i=\mathbf{M}_i/M_i$ is the unit vector of magnetic moment. The autocorrelation function is calculated on-the-fly \cite{Chen_PRB_1994} at thermal equilibrium and averaged using multiple statistically independent starting times. 

The time-displaced correlated behaviour between the $i^{\text{th}}$ magnetic moment and the magnetic moments constituting its $1^{\text{st}}$ coordination shell give insights into the magnetic ordering and timescale of its decoherence:
\begin{equation}
c_1(\mathbf{r},\tau)=\frac{1}{8N}\sum_{n}^{N}\bigg\langle \sum^8_{j=1} \mathbf{e}_j(\mathbf{r'},t+\tau)\cdot \mathbf{e}_i(\mathbf{r},t) \bigg\rangle,
\label{eqn:1nrrcf}
\end{equation}
where atom $j$ is in the 1$^{\text{st}}$ nearest neighbour shell of atom $i$. At cryogenic temperatures, a FM monodomain would yield $c_1=1$ whereas a layered AFM domain such as in L1$_0$ phase FeNi would return $c_1=-1$.

In the following, when we consider a spherical precipitate, we measure the averaged correlation functions of magnetic moments displaced from a chosen origin $\mathbf{r}_0$, i.e. the geometric centre of a cluster. We call $c_0(|\mathbf{r}-\mathbf{r}_0|)$ the radially resolved time-displaced spin-spin autocorrelation functions (RRACF), and $c_1(|\mathbf{r}-\mathbf{r}_0|)$ the 1$^{\text{st}}$ nearest neighbour radially resolved time-displaced spin-spin correlation functions (1nRRCF).

\section{Results and discussion}
\subsection{Disordered Solid Solutions} \label{sec:ss}
Our investigation starts from disordered solid solutions of Fe$_{1-x}$Cr$_{x}$, where $ 0 \leq x \leq 0.25$. We study the change of Curie temeprature $T_C$ through the change of  magnetic susceptibility $\chi$ as functions of Cr concentration and temperature.

SD simulations were performed using a cubic simulation cell with periodic boundary conditions containing 31250 atoms. Collinear ordering of magnetic moments along [001] was chosen for the initial configuration. We define the magnetisation axis parallel to the $z$ direction. Statistics were collected over a 5ps production run following a 1ps thermal equilibration period. Each configuration was generated by random substitution of Cr atoms into the BCC Fe matrix to satisfy the concentration requirements. As we use a relatively large simulation cell, we assume sufficient statistical sampling of local structure. To check this assumption, for $x=15$ and 25 at.\%, we simulated three random configurations and found comparable results for each case.

\begin{figure}
\includegraphics[width=8.5cm]{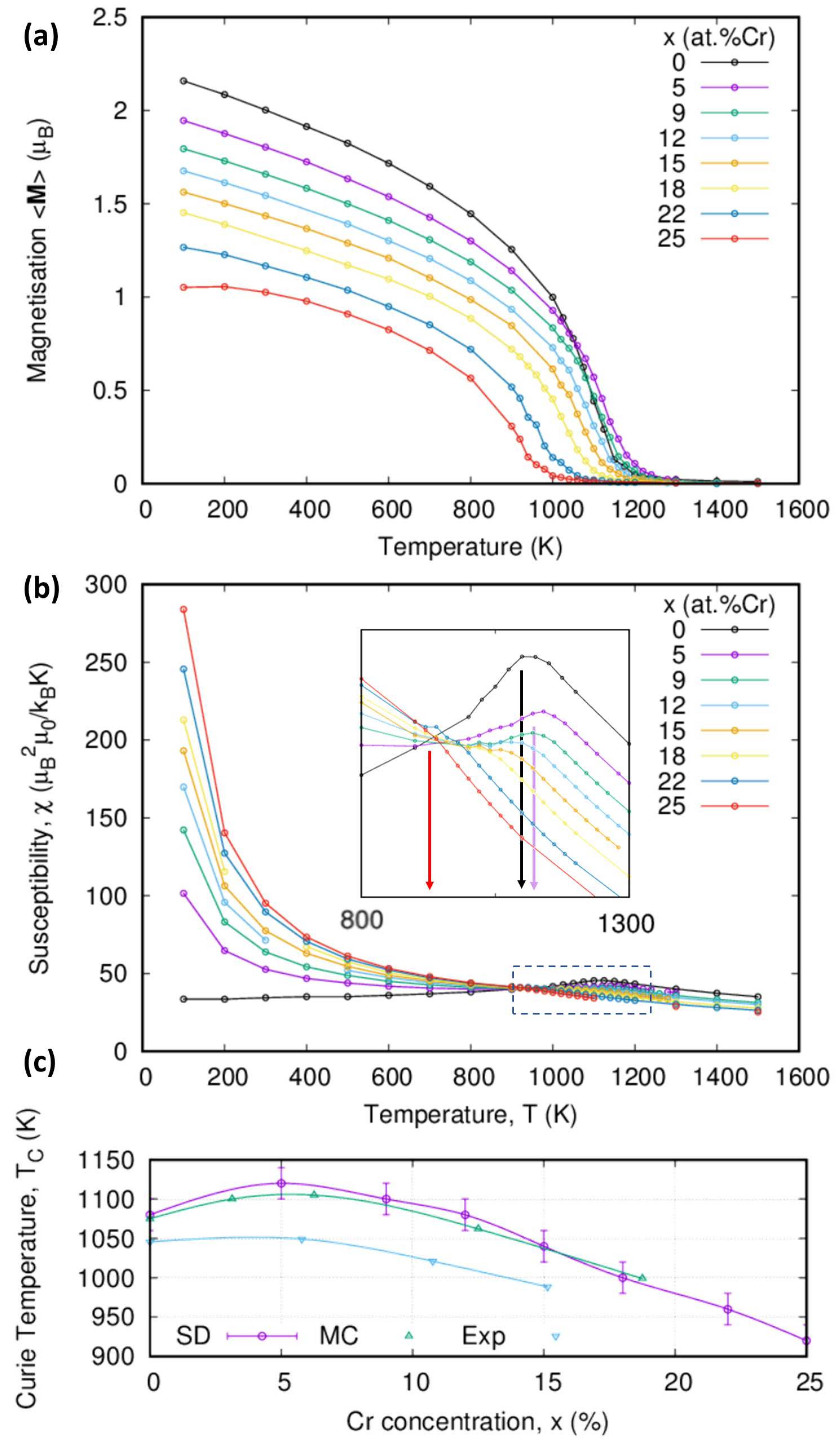}
\caption{\label{fig:tc_ss} Magnetic properties of disordered Fe$_{1-x}$Cr$_{x}$ solid solutions with different Cr concentrations. (a) Magnetisation versus temperature. (b) Magnetic susceptibility versus temperature. (Inset: Susceptibility around the Curie temperature $T_C$.) The points of inflection for $x=0$, 5 and 25 at.\% Cr are indicated by arrows. (c) Curie temperature $T_C$ versus Cr concentration as determined from the susceptibility. Data from Monte Carlo simulations and experiment\cite{Lavrentiev_JPhysCM_2012} are given for comparison. The determination of $T_C$ presented in Ref. \onlinecite{Lavrentiev_JPhysCM_2012} is via observation of peak in the specific heat. Error bars are set as the temperature interval in sampling, which is 20K.}
\end{figure}

The magnetisation of the disordered solid solutions are shown in Fig. \ref{fig:tc_ss}(a) over a temperature range of 100K $\leq T \leq$ 1500K. At low temperatures, the magnetisation decreases as the Cr concentration increases. The magnetisation profiles are similar to pure Fe as the temperature is raised. For Cr concentrations of 5 and 9 at.\%, the magnetisation is found to remain finite to temperatures higher than the $T_C$ of pure Fe. 

We determine $T_C$ through the turning point in the bulk susceptibility ($\partial^2_T \chi = 0$) as shown in Fig. \ref{fig:tc_ss}(b). $T_C$ is plotted against the Cr concentration in Fig. \ref{fig:tc_ss}(c). We note the consolute temperature is approximately 900K. Therefore, disordered solid solutions are appropriate stable configurations near the $T_C$ for aged Fe-Cr alloys across the full compositional range. Data from experiments and MC simulations \cite{Lavrentiev_JPhysCM_2012} are also shown for comparison. In the MC simulations, \citeauthor{Lavrentiev_JPhysCM_2012} \cite{Lavrentiev_JPhysCM_2012} determined the $T_C$ through the observation of peaks in the specific heat instead. Excellent agreement is found between our SD calculations and the MC results, where both show the same trend as experiments. 

$T_C$ increases for small Cr concentrations up to 5 at.\%Cr. Further increases in the Cr concentration decrease the $T_C$. This is in agreement with experimental observations that $T_C$ is maximised for a Cr concentration around 6 at.\% \cite{Lavrentiev_JPhysCM_2012,Xiong_calphad_2011}. The physical mechanism for this phenomenon is known to occur due to the strengthening of the Fe magnetic moment induced by the strong antiferromagnetic coupling to Cr atoms on nearest-neighbour lattice sites \cite{Lavrentiev_JPhysCM_2012,Lavrentiev_PRB_2010,Klaver_PRB_2006}. If an Fe atom has multiple Cr nearest neighbours, it results in magnetic frustration, because antiparallel ordering with respect to the Fe atom cannot be held for all Cr neighbouring moments simultaneously. 

A concentration of $x=6.25$ at.\% Cr is an idealised case. It allows for the highest concentration of Cr which support perfect AFM ordering of all Cr magnetic moments with respect to their neighbouring Fe atoms. This is achieved when the Cr atoms take the corner sites of a repeating unit constructed of $2\times2\times2$ unit cells. This is reflected with a pseudo-gap minimum in the spin-resolved density of states \cite{Nguyen-Manh_ComputMaterSci_2010}.


\subsection{Fe-Cr Alloys with Idealised Cr Clusters} \label{sec:ideal_cluster}

Irradiation alters the microstructure of steels, which influences their mechanical properties \cite{Hardie_JNucMat_2013}. We expect different microstructures to exhibit different magnetic properties, even when the nominal chemical composition of the overall system remains the same. If verified it suggests magnetic properties may be probed externally without requiring the destructive extraction of test samples from components. This may help develop a non-invasive means of assessing microstructure in steels. Towards this end, in this section we investigate how the formation of pure Cr $\alpha'$ precipitates might change the magnetic properties with respect to the disordered solid solutions studied in previous section. 

Atom probe experiments show precipitates form during ageing and irradiation. They starts forming as elongated coagulated clusters and becomes roughly spherical over time \cite{Novy_JNucMat_2009}. We model spherical $\alpha'$ precipitates by substituting Fe atoms with Cr atoms within a sphere of radius $r_c=|\mathbf{r}-\mathbf{r}_0|$. Simulation cells containing 31,250 atoms in BCC structures are created. To achieve a particular nominal composition for the Fe-Cr alloy, we distribute extra Cr substitutions randomly throughout the remaining Fe matrix. 

The system size corresponds to an $\alpha'$ precipitate number density of 2.7$\times 10^{23}$~m$^{-3}$ which is comparable to the lower limits observed experimentally in neutron irradiated Fe-Cr steels \cite{Reese_JNucMat_2018}. Clusters are generated with radii from 6 to 26{\AA} for nominal concentrations of 9, 12, 15, 18, 22 and 25 at.\%Cr. The cluster sizes are chosen to be comparable to precipitates aged from 10 to 10,000 hours \cite{Novy_JNucMat_2009,Reese_JNucMat_2018}. 

\begin{figure}
\includegraphics[width=8.5cm]{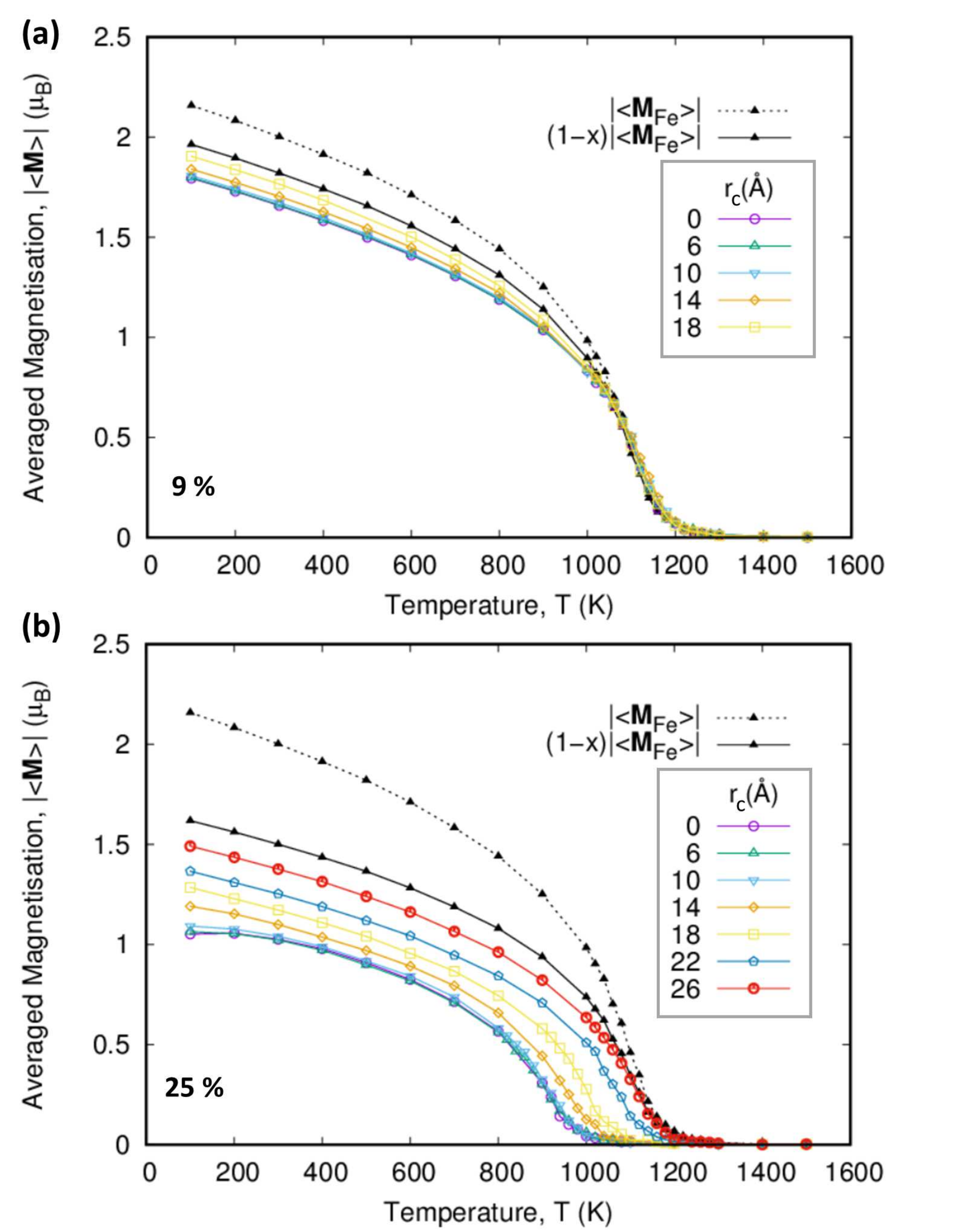}
\caption{\label{fig:mvt} The averaged magnetisation of Fe$_{1-x}$Cr$_{x}$ alloys is plotted as a function of temperature and cluster size with different chemical composition. In each simulation cell, we created a pure Cr $\alpha'$ spherical precipitate. Extra Cr atoms is randomly distributed in Fe matrix to attain the required overall Cr concentration $x$. The radius of precipitates $r_c$ ranges from 6 to 26\AA. Disordered solid solution is denoted as $r_c = 0$\AA. Only $x=$ 9\% and 25\% are shown here. Figures of $x=$12\%, 15\%, 18\% and 22\% are presented in Supplementary Material.}
\end{figure}

\begin{figure*}
\includegraphics[width=17cm]{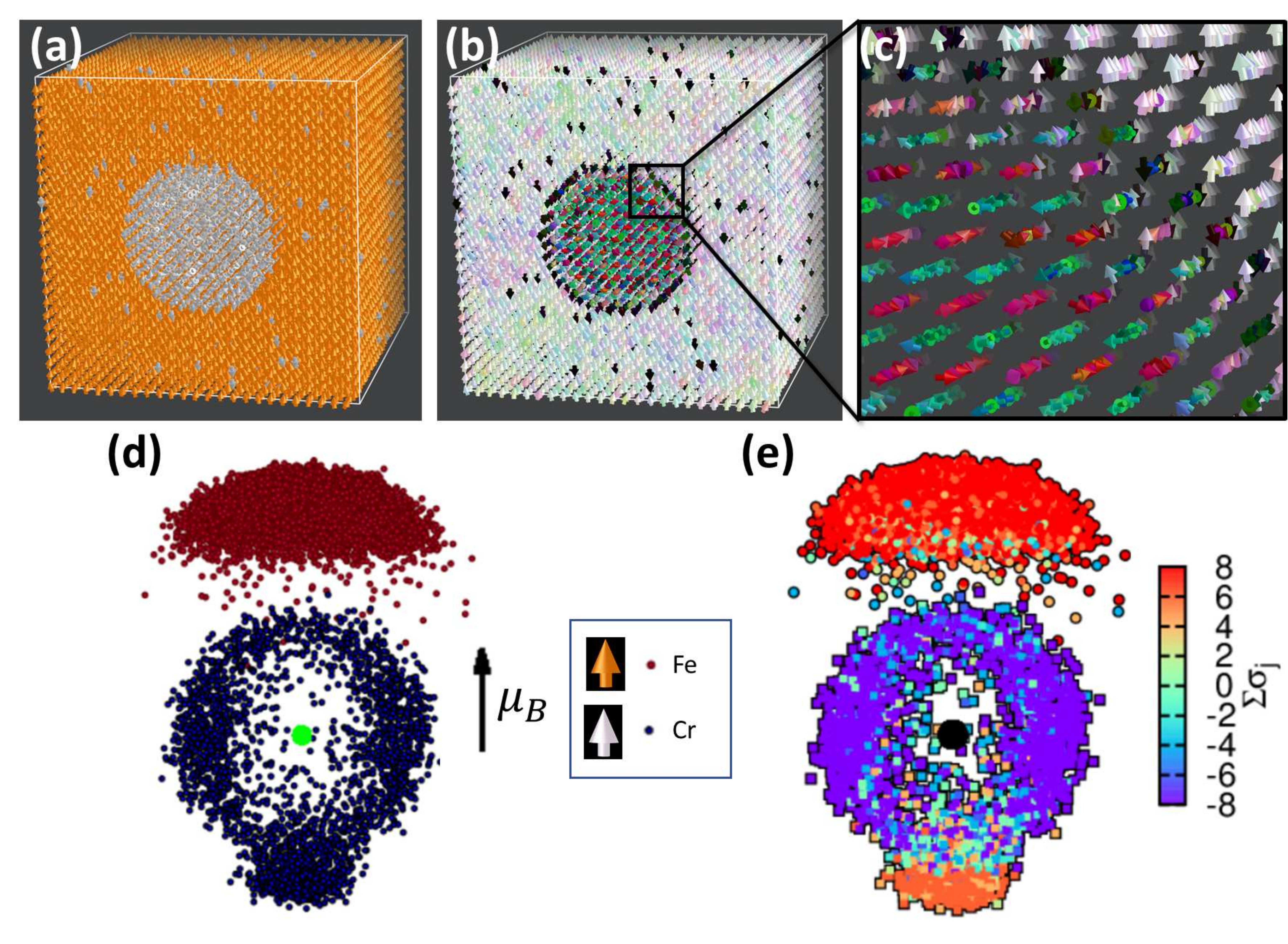}
\caption{\label{fig:cluster_9pc} Snapshot of magnetic moments of a system containing a spherical pure Cr precipitate with radius of 18{\AA}. The simulation cell contains 31,250 atoms with 9 at.\% Cr. The magnetic system is thermalized to 100K. (a) Magnetic moments of Fe is in orange, where Cr is in white. (b) Magnetic moments colourised according to double conical space. (c) Focused view near the interface between precipitate and bulk. (d) Magnetic moment space map of all moments. (e) Magnetic moment space map colourised according to the sum of the occupational site variables of the nearest neighbours, e.g. $\sum \sigma_j=8$ means all neighbours are Fe, 0 means 50\% is Fe and 50\% is Cr, and -8 means all are Cr).}
\end{figure*}

The averaged magnetisation for Cr concentrations of 9\% and 25\%, with different precipitate sizes and at different temperatures, are shown in Fig. \ref{fig:mvt}. Figures regarding Cr concentrations of 12\%, 15\%, 18\% and 22\% are presented in Supplementary Material for completeness. The magnetisation of bulk Fe ($|\langle \mathbf{M}_{\text{Fe}} \rangle|$), the scaled bulk value proportional to the nominal concentration ($(1-x)|\langle \mathbf{M}_{\text{Fe}} \rangle|$) and the magnetisation of the disordered solid solution ($r_c=0$) are also plotted. The scaled bulk Fe curve gives an approximate limit obtained upon complete phase segregation as could theoretically occur within the spinodal decomposition.

In general, the disordered solid solutions have the smallest magnetisation. Both the Fe and Cr moments prefer being antiparallel to neighbouring Cr moments. With disordered configurations (when $x>0.06$) this can rarely be satisfied. In order to minimize the energy, Cr moments try to be as antiparallel as they can with respect to their neighbours, so they tilt. This results in geometric magnetic frustration and non-collinearity of the Cr magnetic moments (see supporting Figure 2(e) in the Supplementary Material). Such ordering reduces the total magnetisation. On the other hand, for the same nominal concentration $x$, the averaged magnetisation increases as Cr is grouped into larger cluster sizes. We find that local atomic configurations with clustering reduces the overall frustration in the system. This reinforces collinearity and enables the constructive superposition of the local moments.
 
In Fig. \ref{fig:cluster_9pc}(a)-(c) we present a snapshot of magnetic moments of a system with a precipitate of radius 18\AA. The Fe-9Cr system has 31,250 atoms. Within the $\alpha'$ precipitate, Cr moments are antiferromagnetic and perpendicular to Fe moments in the $\alpha$ phase. Cr moments situated on the cluster interface are aligned antiparallel to the adjacent Fe moments. They gradually rotate to be perpendicular with the Fe moments when they are well separated from the interface by 5-8\AA. Throughout this tilting from the Fr moments, the Cr magnetic moments attempt to maintain an antiparallel ordering to their Cr neighbours. 

An alternative representation is the magnetic moment space as shown in Fig. \ref{fig:cluster_9pc}(d) and (e). Each dot represents a magnetic moment vector, where all moment vectors are centred at the origin. In (d) the points are coloured according to the elemental species. In (e), the colour corresponds to the number of Fe atoms in the first coordination shell of the atom producing the magnetic moment. Red represents 8 Fe nearest neighbours whereas blue represents 8 Cr. Again, Fe moments in the $\alpha$ are observed to be ferromagnetic. Fe magnetic moments on the interface have less deviance from the magnetisation axis. The antiferromagnetism of the Cr atoms in the $\alpha'$ is shown by the clustering points on the left and right of the moment space representation with a void in the centre. The arcing of these points show the moments which are tilted in the vicinity of the interface. The bulge at the base reveals the magnitude of Cr magnetic moments are strengthened when surrounded by Fe atoms, i.e. those in the $\alpha$ phase.

Non-collinear arrangements of Cr and Fe moments across the interface of the precipitate are similar to observations of (011) Fe-Cr interfaces \cite{Lavrentiev_JPhysCM_2012,Fritzsche_prb_2002}. Our simulations on clusters show that these magnetic configurations are quite general. They relax for low and high index interfaces, and at elevated temperatures, as a method of minimising free energy subjected to magnetic frustration. This highlights the need to consider non-collinear magnetism in the study of steels. 

Knowledge that we just obtained can be applied to realistic materials such as EUROFER-97. This is the European reference steel for the first wall of a DEMO fusion reactor which has a Cr concentration of 9 at.\%. As we can ascertain from Fig. \ref{fig:mvt}, the change of magnetisation is delimited by curves representing the disordered solid solution (0\%) and complete phase separation ($(1-x)|<M_{\text{Fe}}>|$). 

The first wall operating temperature of a fusion reactor is expected to be in the range of 350$^{\circ}$C to 550$^{\circ}$C\cite {Gasparotto_FusEngDes_2003} (i.e. 623 to 823K). We calculated the magnetisation for different Cr concentrations at 600K and 800K. Details can be found in the Supplementary Material. The difference between the magnetisation of the solid solution and complete phase separation at 9 at.\%Cr is 10.4\% at 600K and 10.3\% at 800K. Our analysis shows that there are measurable changes to the macroscopic magnetic properties, which arise due to Cr clustering. 

\begin{figure}
\includegraphics[width=8.5cm]{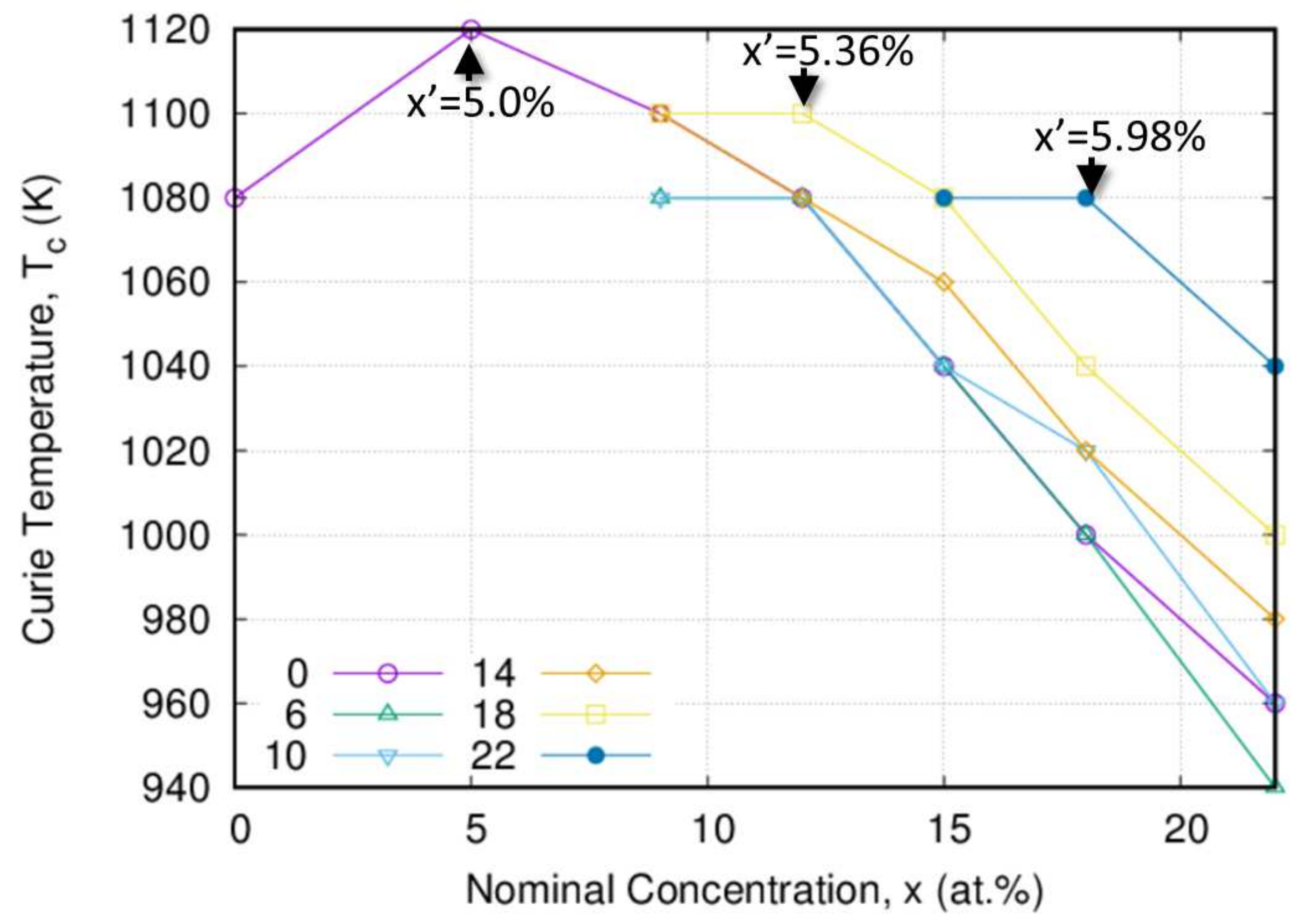}
\caption{\label{fig:Tcvx} The Curie temperature versus the nominal Cr concentration for different $\alpha'$ precipitate sizes with radius of 0, 6, 10, 14, 18 and 22{\AA}. The Curie temperature is maximised when the effective Cr concentration in the Fe matrix is about  5-6 at.\%.}
\end{figure}

In Fig. \ref{fig:Tcvx}, we plot the $T_C$ as a function of nominal composition with different cluster sizes. $T_C$ is measured by the turning point in the magnetic susceptibility. Our calculations show that the $T_C$ of Fe$_{1-x}$Cr$_x$ can vary significantly for a given nominal concentration $x$. For $x=18$at.\% Cr, we find an 80K difference in the $T_C$ between Cr dispersed randomly compared to cases where most atoms are in a large cluster. 

As a metric to quantify the conditions which maximise the Curie temperature, we consider the effective concentration $x'$. This is defined as the concentration of the Cr remaining the $\alpha$ phase (see supplementary material for further details). Indeed for each concentration, we identify that $T_C$ is maximised when enough Cr conglomerates to produce an effective concentration of approximately 6 at.\%Cr. The effective concentration $x'$ in the systems with nominal concentrations $x$ of 5, 12 and 18 at.\%Cr are shown explicitly in Figure~\ref{fig:Tcvx} where $T_C$ is maximised to highlight this trend.

The increase of $T_C$ when $x'\approx 6$ at.\%Cr is not coincidental. It can be attributed to the same mechanism that drives $T_C$ in solid solutions to be maximised at $x=6.25$ at.\% Cr \cite{Nguyen-Manh_ComputMaterSci_2010}. When more Cr atoms are in the precipitate, the relative concentration of Cr within the $\alpha$ phase decreases. This increases the average number of Fe nearest neighbours and decreases the number of nearest and next-nearest neighbouring Cr atoms. This reduces the geometric magnetic frustration of the Cr magnetic moments in the $\alpha$ phase. They can then align antiferromagnetically with the adjacent Fe atoms which is energetically favourable. This in turn strengthens the magnetic moment of the Fe atoms which can remain correlated to higher temperatures. 

Our simulations show the alloy acts as a composite system. The $\alpha$ phase, with an effective concentration of $x'$, behaves as a disordered solid solution Fe$_{1-x'}$Cr$_{x'}$. Cr within the $\alpha'$ precipitate are bulk-like with an antiferromagnetic ordering below the N\'{e}el temperature. In the succeeding sections we consider the dynamics at the cluster interface and how Fe substitutions within the $\alpha'$ change the magnetic properties.

\subsection{Spatial and Time Resolved Magnetic Properties of $\alpha'$ Clusters}\label{ss:time}

\begin{figure}
\includegraphics[width=8.5cm]{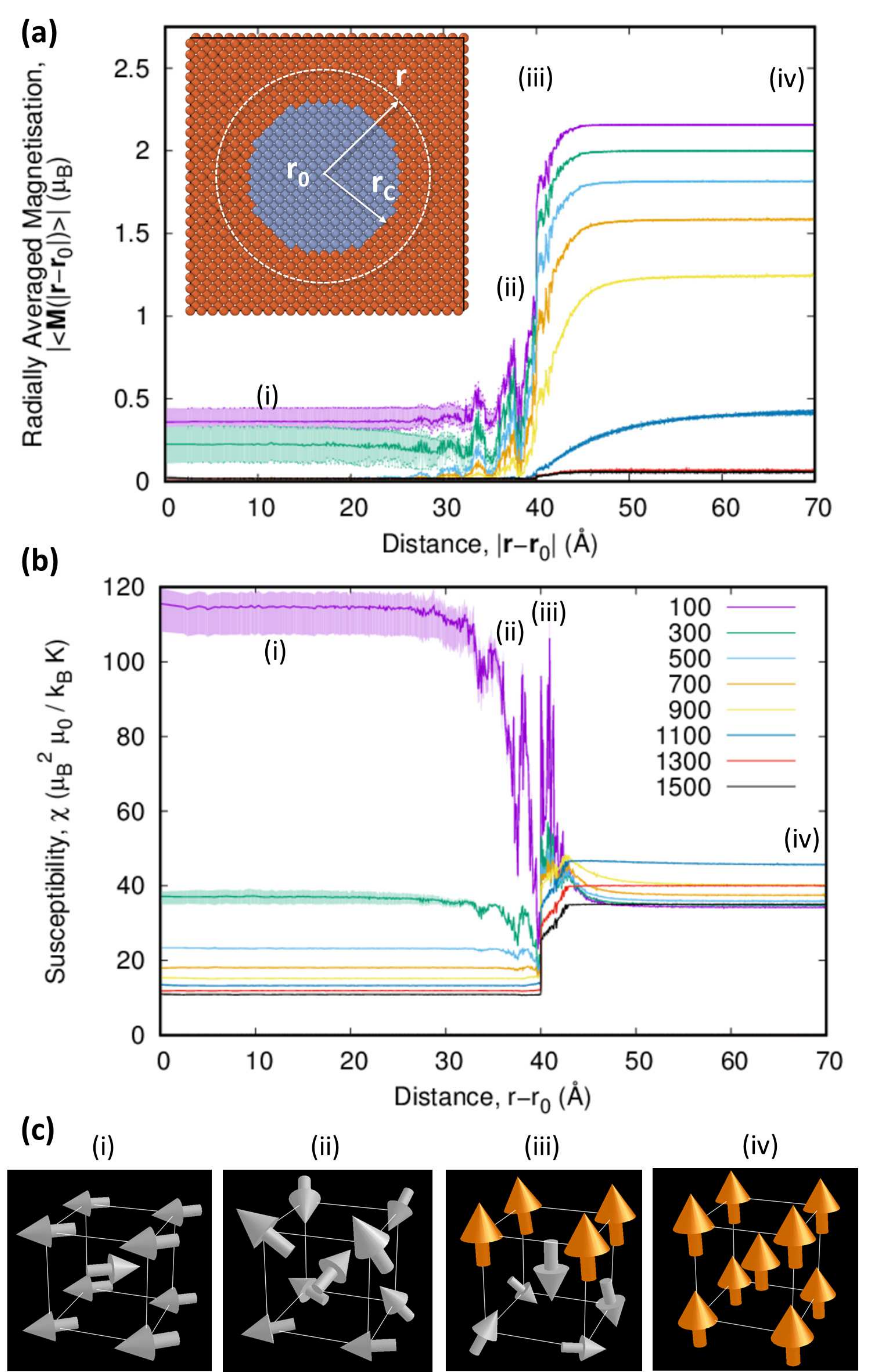}
\caption{\label{fig:xc1} Magnetic properties of a spherical pure Cr precipitate with radius $r_c=4$~nm in pure Fe. (a) Radially averaged magnetisation and (b) susceptibility evaluated as a function of distance $|\mathbf{r}-\mathbf{r_0}|$ from the centre of the precipitate. (c) Time averaged magnetic configurations for $|\mathbf{r}-\mathbf{r}_0|$ = (i) 10.1{\AA}, (ii) 35.1{\AA}, (iii) 40.0{\AA} and (iv) 69.97{\AA} at $T=$100K.}
\end{figure}

\begin{figure*}
\includegraphics[width=17cm]{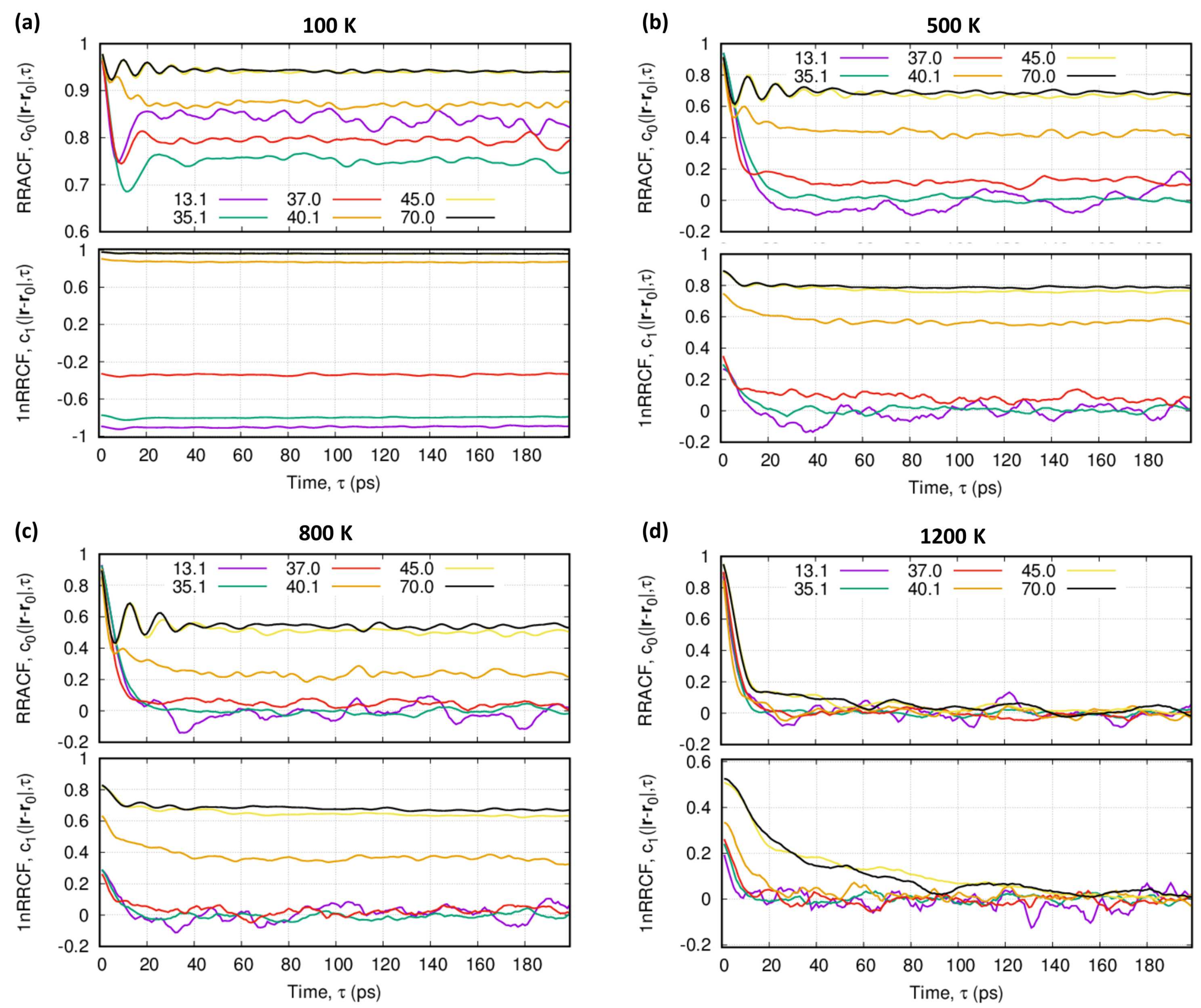}
\caption{\label{fig:id_corr} Radially resolved autocorrelation and first nearest neighbour correlation functions at $T=$100K, 500K, 800K and 1200K. They are taken as averaged values for magnetic moments located at 13.1, 35.1, 37.0, 40.1, 45.0 and 70.0{\AA} away from the centre of a Cr precipitate in pure Fe with a  radius of 40{\AA}.}
\end{figure*}

For larger cluster sizes, we can study the magnetic properties with spatial resolution. We demonstrate this by examining a spherical Cr cluster, of radius 4~nm with no impurities, situated within pure Fe. We used a simulation cell containing 250,000 atoms, that is $50\times 50 \times 50$ unit cells. This corresponds to a number density of 3.2$\times 10^{23}$~m$^{-3}$ which can be observed experimentally, such as experiments on neutron irradiated Fe-12Cr at 320$^{\circ}$C by \citeauthor{Reese_JNucMat_2018} \cite{Reese_JNucMat_2018}.

In Fig. \ref{fig:xc1} the radially averaged magnetisation and susceptibility for temperatures from 100K to 1500K are plotted as a function of distance from the centre of the cluster $|\mathbf{r}-\mathbf{r}_0|$. The magnetisation in the pure Fe surrounding a Cr cluster are found to have bulk values. When approaching the Cr-Fe interface from the $\alpha$ phase, the magnetisation drops when the Fe coordination shells become populated by Cr atoms. The magnetisation drops sharply within the $\alpha'$. Both the susceptibility and averaged magnetic moment are observed to oscillate within a penetration depth of 1 to 2nm in the $\alpha'$. These behaviours are due to the long range nature of the exchange coupling parameters. 

We repeated the calculation three times with slightly different initial magnitudes of the magnetic moments such that different phase space trajectories were explored. We did this in order to confirm the oscillations are physical, and not caused by noise or poor convergence. It also improves our statistical sampling. The average is plotted as the solid line for each temperature with the shaded region limiting the maximum and minimum values of the repeated runs. We confirm the oscillations at $32${\AA} $ < |\mathbf{r}-\mathbf{r}_0| < 40${\AA} are physical by noting only small variances between the peaks of all the calculations. Furthermore, the peaks occur at fixed distances across different temperatures.

To visualise the local magnetic configurations, the time averaged magnetic moments within a representative conventional unit cell centred at each radius is shown in Fig. \ref{fig:xc1}(c). Four radii of interest were selected (i) 10.1{\AA} (ii) 35.1{\AA} (iii) 40.0{\AA} and (iv) 69.97{\AA}. Cr magnetic moments situated deep within the cluster are arranged antiferromagnetically. They are perpendicular to the ferromagnetic Fe moments in the $\alpha$. The magnitudes of Cr moments on the interface are enhanced. At the interface their orientations are locked antiparallel to the interfacial Fe moments. Cr atoms further away from the interface rotated in plane. A region of frustration exists within a penetration depth of 10{\AA} from the interface were the Cr moments are less correlated. These properties agree with the smaller scale calculations in the previous section.

The susceptibility within a cluster is particularly high when the temperature is below the N\'{e}el temperature $T_N\approx 350$K. This is due to the magnetic frustration of the Cr near the interface. A peak in susceptibility is observed within the Fe matrix even at a temperatures much lower than $T_C$. From Eq. \ref{eqn:sus}, it is apparent that the Fe moments near the interface fluctuate more readily than in the bulk Fe. We can deduce that the disorder of the configuration at the Fe-Cr interface leads to a reduction of $T_C$ locally. This is evidenced further in the analysis of correlation functions below.

We note that the magnetisation level in the centre of a pure Cr cluster is not unique. This indicates there is metastability enabling the magnetic moments to order according to initial conditions akin to a spin glass. The system also meets the criteria of frustration due to the local configuration dependence of the exchange parameters. Since Fe-Cr is known to become a spin glass for 81-84\%Cr \cite{Burke_JPF_1983}, it would be interesting to determine if Cr rich $\alpha'$ precipitates form spin glasses locally for lower nominal concentrations.

%
%
An advantage of SD compared to MC is its capability of simulating the dynamic time resolved magnetic properties. In Fig. \ref{fig:id_corr}, we plot the radially resolved autocorrelation function (RRACF, Eq. \ref{eqn:rracf}) and the first nearest neighbour radially resolved correlation function (1nRRCF, Eq. \ref{eqn:1nrrcf}) of magnetic moments at temperatures of 100K, 500K, 800K and 1200K. They are measured at different distances from the centre of the cluster. We selected six radii to compare. Three of them are within the precipitate, 13.1{\AA}, 35.1{\AA} and 37.0{\AA}. The next is at 40.1{\AA}, which is at the first Fe layer surrounding the Cr cluster. The remaining two are at 45.0{\AA} and 70.0{\AA}, which are in the $\alpha$ phase.

%
%
The profile of the correlation functions $c_0$ and $c_1$ consist of a decaying envelope function indicative of the characteristic relaxation or dephasing time, where its asymptotic value represents the strength of correlation. The oscillations in the RRACF are due to the precession of magnetic moments\cite{Ma_PRB_2008}. Fe moments precess at a higher frequency than the Cr moments. If the rotational frequency $\omega_i = -|\mathbf{H}_i|/\hbar$ is considered, we immediately realise the effective field is greatest on the Fe moments. Correspondingly the effective field is weak where there is reduced correlation such as at 35.1\AA.  A relaxation time of 20 to 40ps is generally observed.

%
%
At $T=100$K the RRACF is positive for all radii. In the $\alpha$ phase away from the cluster, the RRACF remains close to unity indicating the magnetic moments are strongly correlated over time. The Fe moments on the interface (40.1{\AA}) have a lower RRACF due to the magnetic frustration reducing the correlation. This coincides with the sharp increase in magnetic susceptibility along the interfacial Fe layers. Near the centre of the precipitate (13.1{\AA}), the Cr magnetic moments have bulk-like ordering and are strongly correlated. 

%
%
Near the interface the magnetisation within the cluster fluctuates. As one can see in Fig. \ref{fig:xc1}(a), the radially averaged magnetisation peaks at 37.0{\AA} and troughs at 35.1\AA. The RRACF shows that at 100K, the moments are more correlated at 37{\AA} where a peak is observed in the magnetic moment (see Fig.~\ref{fig:xc1}a). This contrasts with 35.1~{\AA}~which corresponds to a radius of reduced magnetisation. Further for 35.1\AA, the RRACF is weakest indicating localised frustration. 

%
%
The 1nRRCF provides a measure of correlation between a magnetic moment at $|\mathbf{r}-\mathbf{r}_0|$ with its nearest neighbours. At 100K, it shows that the Fe moments in the $\alpha$ phase remain extremely collinear with their neighbours since $c_1\approx 1$. Collinearity is decreased slightly for the Fe magnetic moments near the Cr cluster (40.1\AA). Cr magnetic order is bulk-like near the centre of the precipitate at 13.1{\AA} where $c_1 \approx -1$. Towards the Fe interface the Cr moments become less correlated with their nearest neighbours over time as indicated by the increased value of the 1nRRCF. This indicates the Cr moments are deviating from AFM ordering. Indeed this assessment agrees with Fig. \ref{fig:xc1}c(ii) where magnetic moments are seen to be highly non-collinear with their neighbours.

%
%
At 500K, Cr moments in the centre of the precipitate (13.1{\AA} and 35.1{\AA}) become uncorrelated, which is consistent with $T_N\approx 350$K. These moments dephase within 20-30ps. Moments at 37.0{\AA} from the precipitate centre are still weakly correlated with their neighbouring atoms, and the 1nRRCF becomes positive. This implies the local $T_N$ is enhanced for Cr near the interface in contrast to the reduction in $T_C$ for interfacial Fe. 

%
%
At 800K, Fe magnetic moments remain strongly correlated as shown in RRACF and 1nRRCF albeit collinearity has decreased due to thermal fluctuations. Fe moments near the $\alpha'$ interface at 40.1{\AA} remain less correlated than those in the Fe matrix and take a longer time to dephase. These interfacial Fe moments have a higher oscillation frequency indicating that their effective fields are stronger because of the strong $J_{1nn}^{Fe-Cr}$ interaction. The larger fluctuations give rise to the enhanced susceptibility (Eq.~\ref{eqn:sus}). This provides further evidence that the effective $T_C$ is reduced around the cluster. Consequently the interface acts as a nucleation site for the ferromagnetic to paramagnetic phase transition.

%
%
At 1200K, all magnetic moments become incoherent, as shown by both the RRACF and the 1nRRCF dropping to zero (Figure~\ref{fig:id_corr}c). This is consistent with the loss of the net magnetisation (Fig.~\ref{fig:xc1}a). The dephasing time of the bulk Fe moments with their neighbours is approximately ten times greater than Cr moments within the $\alpha'$ cluster.   

%
%

\subsection{Impure Cr Clusters}\label{sec:clusters}
\begin{figure*}
\includegraphics[width=17cm]{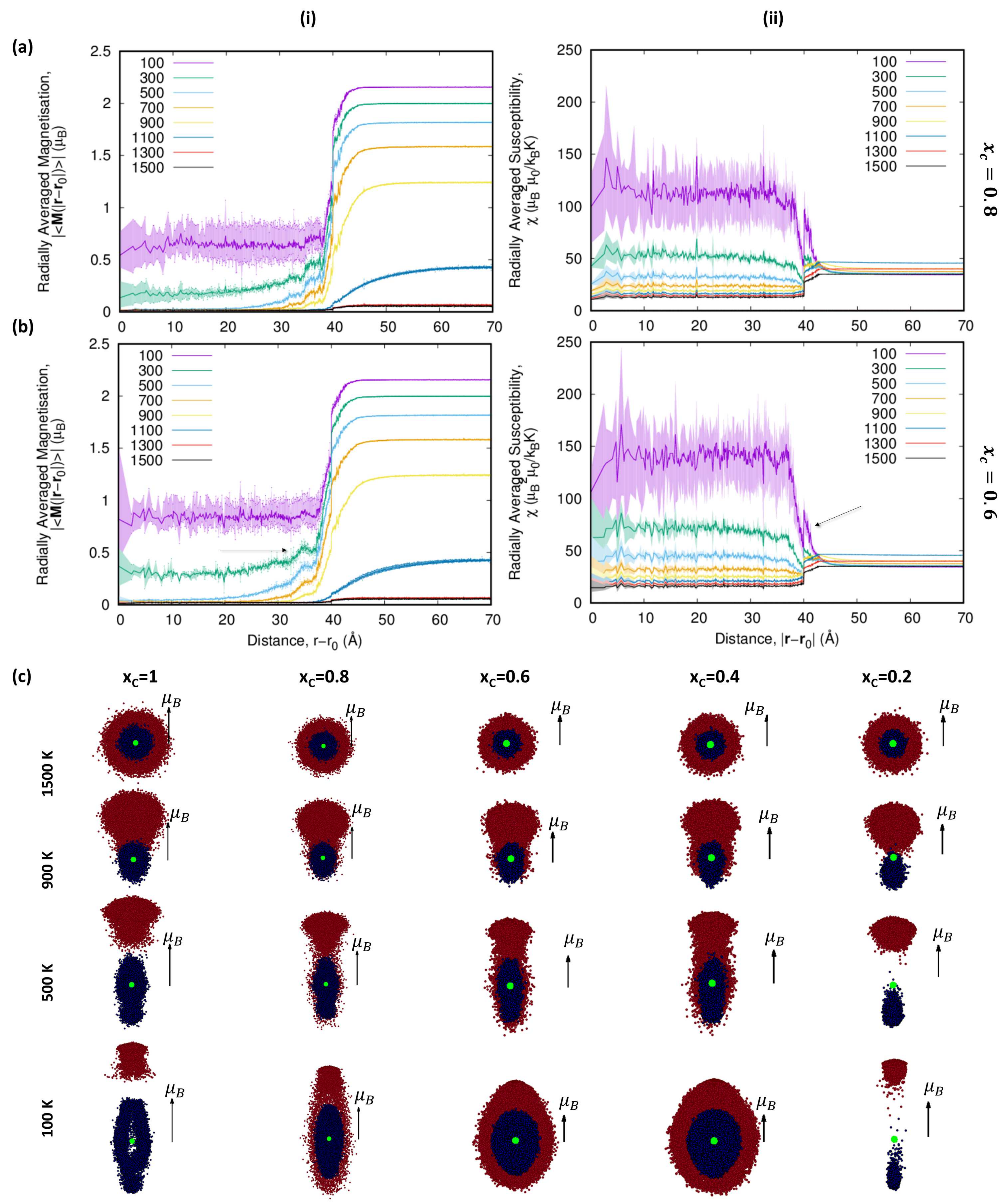}
\caption{\label{fig:xc_mxvt} (i) Radially averaged magnetisation and (ii) magnetic susceptibility as a function of radius from the centre of a spherical $\alpha'$ cluster with radius of 40\AA. The Cr precipitate has a concentration of (a) 80 at.\%Cr, (b) 60 at.\%Cr. Concentrations of 40 at.\%Cr and 20 at.\%Cr are provided in Supplementary Material. Three samples are simulated for each case. The averaged value is shown as the solid line. The region between the smallest and largest values measured for each simulation is shaded. (c) Time averaged magnetic moment space diagrams for the Cr clusters of different local concentrations at 100K, 500K, 900K and 1500K. Each dot represent a magnetic moment centred at the origin. Fe moments are in red, where Cr moments are in blue. Each map is shown with respect to a scale of 1$\mu_B$. Magnetic moments were averaged over every time step during the 5~ps production run.}
\end{figure*}

In previous sections, we considered idealised spherical $\alpha'$ clusters. In reality, a Cr cluster is rich in Cr but impure \cite{Reese_JNucMat_2018}. Recent studies show Cr clusters formed in neutron, ion and electron irradiated alloys, as well as in thermally aged materials, have sizes between 10 and 40{\AA} and have local Cr concentrations between 50 and 95\% \cite{Tissot_ScriptMater_2016,Rogozhkin_InorgMaterApplRes_2016,Chen_JNucMater_2015,Reese_JNucMat_2018}. We investigate whether the same phenomena exhibited by pure Cr clusters can also be observed with impure clusters as the $\alpha'$ phase is diluted by Fe substitutions. 

Spherical $\alpha'$ clusters with Cr concentration $x_c$ and radius 40{\AA} are created in a pure Fe matrix. Three configurations are generated for concentrations of $x_c=0.8,0.6,0.4,0.2$ in a supercell with 250,000 atoms. Since irradiated steels have been experimentally observed to have $\alpha'$ clusters with concentrations between 50 and 95\% we report the $x_c=0.8$ and 0.6 cases in Fig. \ref{fig:xc_mxvt}. Figures for $x_c=$0.4 and 0.2 are in Supplementary Material. In Fig. \ref{fig:xc_mxvt}, column (i) shows the radially averaged magnetisation and column (ii) the magnetic susceptibility as a function of distance from the centre of the cluster. The region between the smallest and largest values measured for each simulation is shaded and the average of the three runs is plotted as a solid line.

At low temperatures ($T<T_C$), the magnetisation in the Fe matrix recovers the bulk values typically within 10{\AA} from the $\alpha'$ interface. When the temperature increases, the recovery length increases for all $x_c$. At 1100K, the recovery length is 20-30{\AA} for all $x_c$, even where there is no Cr in the Fe matrix. This indicates the interactions and associated reduced ferromagnetic ordering proximate to the interface have long range effects, which reduces their correlated reinforcement.

For higher concentrations of Fe within the $\alpha'$, the total magnetisation within the cluster steadily increases. A discontinuous change in the radially averaged magnetisation is observed across the interface. The total magnetisation is enhanced within 10{\AA} of the precipitate interface.

The enhanced magnetic susceptibility of the Fe atoms near the interface discussed in the previous section is also observed for the non-ideal clusters (see arrow in Fig.\ref{fig:xc_mxvt}(b)(ii)). When the local Cr concentration decreases, the enhancement decreases. Nonetheless, even for the $x_c=0.2$ case the susceptibility of Fe on the interface is still increased relative to the rest of the $\alpha$ phase. This indicates that $\alpha'$ acts as a nucleation site for the phase transitions even for clusters with dilute Cr concentration. 

Within the cluster, the susceptibility is reduced near the interface. This corresponds to an increased effective $T_N$ originating from the interfacial Cr moments coupling to Fe moments.

To visualise the magnetic ordering in the system and the change with temperature, the magnetic moment space for each precipitate concentration is shown in Fig. \ref{fig:xc_mxvt}(c). The moments are averaged over every timestep during the 5~ps SD simulations.  For $x_c=1$ the Cr magnetic moments in the centre of the cluster arrange in antiferromagnetic order and perpendicular to the magnetisation axis of the Fe as was shown in Fig. \ref{fig:cluster_9pc}. As the temperature increases, the moments became more disordered. Nonetheless, at 500K, which is above $T_N$, Cr moments retain some ordering and remain mostly parallel to the Fe magnetisation axis as indicated by the prolated distribution. At 1500K, when temperature is well above $T_N$ and $T_C$, the moments of Fe and Cr are spherically distributed indicating a paramagnetic state. 

For $x_c=0.8$, 0.6 and 0.4, the Fe moments dispersed within the $\alpha'$ precipitate experience significant frustration. This non-collinearity is shown in the magnetic moment space by the egg-shaped / ovoid distribution. The Fe moments maintain weak ordering with a net magnetisation and with maxima along the major axis of the ovoid. The Cr moments are similarly distributed albeit with a weak total magnetic moment orientated opposite to that of Fe moments. Surprisingly at 500K, the collinearity increases. The thermal agitation enables Fe moments perpendicular to the magnetisation axis to reorient. We note that experiments have confirmed spin glass state 81-84 at.\% Cr at low temperatures \cite{Burke_JPF_1983}. This spin glass state transits into ferromagnetic state as temperature increases. Our results suggest Cr clusters share glassy characteristics.

In summary, the idealised pure Cr precipitates have an ordered non-collinear AFM structure which becomes more disordered with Fe substitutions. Despite the reduction of magnetic ordering, the overall properties remain similar. The effective $T_C$ of Fe surrounding the $\alpha'$ cluster is reduced whereas $T_N$ of interfacial Cr atoms is increased. This results in a spatial dependence of time-resolved properties such as the susceptibility which was used to identify interfacial Fe to be more susceptible than Fe in the $\alpha$ phase.

\subsection{Ideal (001) Interface of Fe-Cr Superlattices}\label{sec:interface}
For large clusters, we may approximate the interface by ideal flat interfaces with different orientations. Indeed, it has been shown that when non-collinear magnetic configurations are permitted, the formation energies of (001), (011) and (111) Fe/Cr interfaces predicted using density functional theory are nearly degenerate \cite{Lavrentiev_PRB_2011}. As such, in this section, we limit our investigation to the study of the magnetic structure of a (001) Fe/Cr interface. 

Spin dynamics calculations are performed on a cell containing 32000 atoms, where $20\times 20\times 20$ Cr unit cells stacked upon $20\times 20\times 20$ Fe unit cells. Periodic boundary conditions are enforced. Essentially, we create an Fe/Cr superlattice with two interfaces in the principle cell. The time averaged magnetic properties of such a confirguration has been studied using MC MCE simulations in Ref. \onlinecite{Lavrentiev_PRB_2011}.  In the Supplementary Material we provide the averaged magnetisation of the superlattice system and the layer resolved magnetic moments which we compare to these earlier calculations by \citeauthor{Lavrentiev_PRB_2011} \cite{Lavrentiev_PRB_2011}. 

The magnetisation almost equals to the scaled value for bulk Fe implying the Fe magnetic moments in the Fe layers remain bulk like at low and intermediate temperatures. It is similar to the case of large clusters with high Cr concentration. At higher temperatures the magnetisation of the interfacial system reduces faster than the bulk Fe (Supplementary Fig. 3). We can infer the $T_C$ is reduced by approximately 100K. From the layer averaged calculations, the magnetisation of the middle Cr layers, the averaged magnetisation vanishes between 300 and 400K corresponding to the N\'{e}el temperature ($T_N\approx 350$K). A finite averaged moment in Cr is observed within a penetration depth of about 20{\AA} for all temperatures $<T_C$. The middle layers of the Fe recover bulk magnetisation values at low temperatures. For Fe in layers close to the interface, the magnetisation is suppressed. 

The bulk magnetisation recovery length from the interface increases with temperature. There is excellent agreement between our SD calculations with the MC simulations for these time averaged properties. Using SD, we can also study the continuous evolution of magnetic moments as a function of time. We analyse the time correlation between spins by considering correlation functions across a simple (001) interface. These results are compared to the dynamics observed for more complex interfaces of clusters. This allows us to conclude whether the behaviour identified in Section \ref{ss:time} is general or is a consequence of the topology of a cluster.

\begin{figure}
\includegraphics[width=8.5cm]{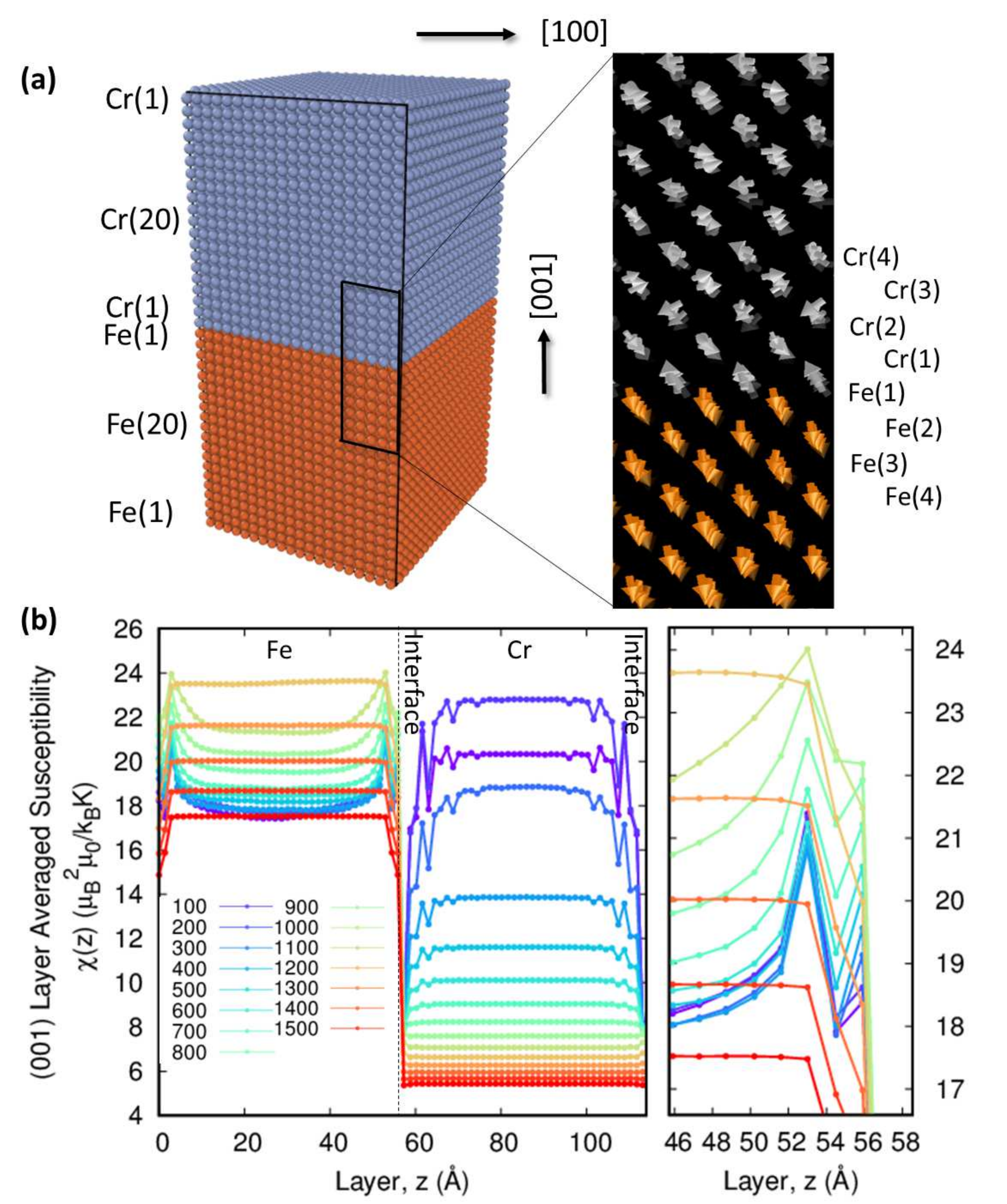}
\caption{\label{fig:interface_X}(a) Schematic view of an Fe/Cr superlattice with (001) interfaces, and a snap-shot of the magnetic moments at the interface at 100K. Fe moments are in orange. Cr moments are in grey. (b) Magnetic susceptibility as a function of temperature for each (001) layer in a Fe/Cr superlattice. }
\end{figure}

In Fig. \ref{fig:interface_X}(a), we show a snapshot of the magnetic moments near the interface at 100K. 
The Fe moments remain collinear whereas the orientation of the Cr moments are strongly dependent upon their proximity to the interface where they have a larger proportion of Fe in their first five coordinate shells. The first interfacial Cr layer is orientated antiparallel to the Fe moment alignment. The deeper into the Cr slab, the magnetic moments reorientate gradually until perpendicular to the Fe slab. Since each Cr sublayer is attempting to minimise energy by aligning antiparallel to one another, there is substantial frustration near the interface. This phenomenon was likewise observed near the Cr cluster interface.

\begin{figure}
\includegraphics[width=7cm]{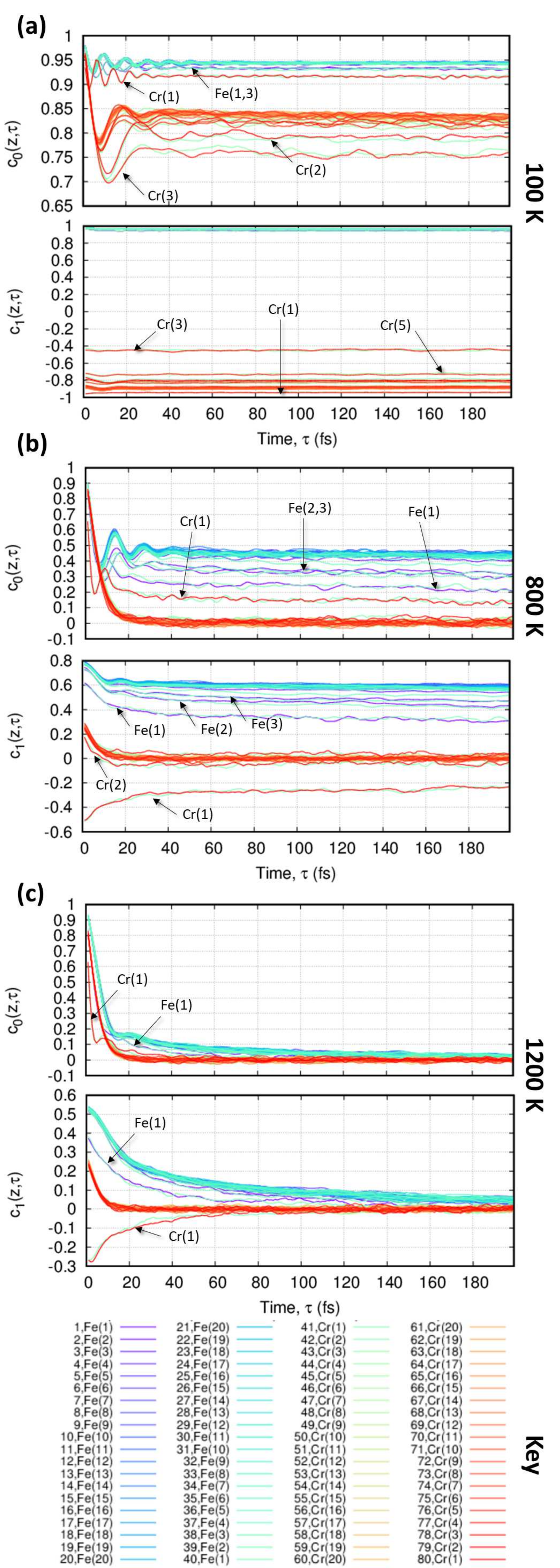}
\caption{\label{fig:interface_corr} Time-displaced spin-spin autocorrelation and nearest neighbour correlation functions of each (001) layer in a Fe/Cr superlattice at 100K, 800K and 1200K. Fe/Cr interfaces are located between layers 80 \& 1, and 40 \& 41. Curves of interest are labelled according to the atoms in the layer and their proximity to the interface. The Fe interfacial layers (1,40) are labelled as Fe(1). The next layers (2,39) are labelled Fe(2), Cr interface layers (41,80) are labelled Cr(1), and so on.}
\end{figure}

In Fig. \ref{fig:interface_X}(b), the susceptibility resolved for each (001) layer in the simulation cell is shown between 100~K and 1500~K with 100~K increments. Remarkable similarity is observed with respect to our results for an ideal $\alpha'$ cluster (Fig. \ref{fig:xc1}(b)). Again, enhanced susceptibility is identified for the Fe moments near the Fe/Cr interface. It becomes clear that presence of Cr destabilises the Fe ordering near an interface irrespective to the index of the plane. 

Additional evidence can be observed from the time correlation functions (Eq. \ref{eqn:rracf} and \ref{eqn:1nrrcf}) which are now defined to be resolved per (001) layer in Figure \ref{fig:interface_corr}. Each layer is labelled by the elemental type occupying the layer (either Fe or Cr) and a number indexing the layers proximity to the interface, such as Fe(1) or Cr(4). Index 1 represents interfacial layers (sample indices are shown in Figure \ref{fig:interface_X}a).

In Fig. \ref{fig:interface_corr}, the autocorrelation function $c_0(z,\tau)$ shows the 1$^{\text{st}}$ and 3$^{\text{rd}}$ Fe layers (Fe(1,3)) at 100K are less correlated than the 2$^{\text{nd}}$ layer. This corresponds to the dip recorded in the susceptibility below 800K (see Fig. \ref{fig:interface_X}b). Above 800K, the 2$^{\text{nd}}$ Fe layer becomes increasingly uncorrelated with respect to the interface layer Fe(1). 

The interfacial Cr layers Cr(1), remain strongly correlated in comparison to the other Cr layers.  A minimum is observed in the susceptibility at the interface for all temperatures correspondingly. At 800K, almost twice $T_N$, $c_1(z,\tau)$ identifies the moments in Cr(1) are still correlated to the moments of the nearest neighbouring atoms. This behaviour of the susceptibility and correlation is observed for both the (001) interface and the high index interfaces of spherical $\alpha'$ clusters. This shows simple (001) interface are good approximations to more complex interfaces. 
\section{Conclusion}\label{sec:conclusion}

Spin dynamics simulations were performed to investigate the magnetic properties in Fe$_{1-x}$Cr$_{x}$ alloys. Different Cr concentrations and degrees of Cr clustering were considered. We adopted a magnetic cluster expansion Hamiltonian that considers both longitudinal and transverse fluctuations of magnetic moments within the framework of Langevin spin dynamics. We observed a strong dependence between the magnetic properties of the FeCr alloys with the local microstructure.

Using the SPILADY code \cite{Spilady} we compared the change of Curie temperature between disordered solutions and systems with precipitation. In both cases the Curie temperature was determined from the magnetic susceptibility. For disordered solid solutions, the Curie temperature was maximised for a Cr concentration of approximately 5 at.\%. When Cr clustering was considered, the total magnetisation for any nominal Cr concentration increased with the size of the Cr precipitate at low temperatures and was minimum for solid solutions.  Configurational disorder in the solid solution created geometric magnetic frustration that reduced the magnetisation. This means that the higher is the level of Cr precipitation, the higher is the total magnetisation.

When we observed the change in the Curie temperature for nominal concentrations of Cr with precipitates of varying size, we found a dependence upon the effective concentration. The effective concentration is the Cr concentration outside the $\alpha'$ cluster. The transition temperature was found to be highest for configurations with an effective $\alpha$ concentration of 5-6~at.\%.

For a particular case of Fe-9Cr alloys, which has the same Cr concentration as EUROFER-97, the magnetisation can vary by 10\% over the operational temperature ranges of the first wall of fusion power plant, for example from 600K to 800K. The change of magnetisation is due to the varying level of Cr precipitation. This potentially has significant impact on the integrity of the mechanical properties of steels, which is coupled to its magnetic properties. 

Cr located at the $\alpha$-$\alpha'$ interface were observed to have moments that remained strongly correlated over time and with respect to the neighbouring moments. The Cr moments were aligned anti-parallel to the moments of adjacent Fe atoms. This reduced their susceptibility relative to Cr in the core of the cluster. 

For pure Cr precipitates, the moments were identified to rotate from being anti-parallel with Fe at the interface, to being perpendicular in the centre. The Cr moments were in an antiferromagnetic configuration with respect to their nearest neighbours. Cr moments near the interface, where the non-collinear configuration was observed, were determined to be the least correlated and therefore frustrated. For impure clusters of experimentally consistent Fe concentrations (50-95\%Cr), the ordered antiferromagnetism in the centre of the cluster was destroyed. Instead, disordered non-collinear configurations were observed at all temperatures.  A net magnetisation was still supported. 

When temperature was increased, the Fe moments close to the interfaces became more disordered than in the bulk. The Cr cluster interface acts as a nucleation site with a reduced effective Curie temperature. Conversely, the Cr moments on the interface remain statistically correlated at temperatures far exceeding the N\'{e}el point.

We may understand the underlying physics through an investigation into dynamic properties which can be calculated from spin dynamics calculations in contrast to the Monte Carlo method. We studied the spin-spin autocorrelation and $1^{\text{st}}$ nearest neighbour correlation functions in Cr clusters and (001) Fe-Cr superlattices. Contrary to expectation, the high index of the cluster interface did not significantly change the observed correlation of Fe and Cr across the interface with respect to the simple (001) case. Fe magnetic moments near the interface fluctuate more readily than in the bulk of the $\alpha$. This resulted in enhanced susceptibility of the Fe in the interfacial layers. The dynamic properties observed for the impure clusters remained consistent with the pure clusters and (001) interface.

The effect of external magnetic field, which was not considered here, may indeed worth attention from the technological perspective. In the design of ITER\cite{international2001iaea}, the maximum toroidal magnetic field will be 12T and poloidal field will be 6T. According to a spin-lattice dynamics study on pure iron\cite{Chiu_aipa_2014}, they may increase the transition temperature by approximately 1K T$^{-1}$. Further verification is required for Fe-Cr alloys. A quantitative analysis of dynamic magnetism as a function of both microstructure and an external magnetic field would be insightful future work for reactor material design.

In summary, using spin dynamics we have investigated the magnetisation, the susceptibility and the Curie temperature of Fe$_{1-x}$Cr$_{x}$ alloys with different Cr concentration and different degrees of Cr precipitation. This enhances our knowledge of how magnetic properties of steels change upon ageing and under irradiation damage. This may help to find a non-invasive way of determining the level of Cr precipitation through the observation of magnetic properties of alloys.  \\

\begin{acknowledgments}
We would like to express our gratitude to Mikhail Yu. Lavrentiev for valuable discussions and providing data from his work. This work has been carried out within the framework of the EUROfusion Consortium and has received funding from the Euratom research and training programme 2014-2018 and 2019-2020 under grant agreement No. 633053 and No. 755039 and from the RCUK Energy Programme [grant number EP/P012450/1]. SLD acknowledges support from the Centre for Non-Linear Studies at LANL. To obtain further information on the data and models underlying this paper please contact PublicationsManager@ukaea.uk. The views and opinions expressed herein do not necessarily reflect those of the European Commission. 
\end{acknowledgments}

\bibliography{main.bbl}

\end{document}